\documentclass[lettersize,journal]{IEEEtran}
\usepackage{amsmath,amsfonts}
\usepackage[spanish,english]{babel}
\usepackage{algorithmic}
\usepackage{array}
\usepackage[caption=false,font=normalsize,labelfont=sf,textfont=sf]{subfig}
\usepackage{textcomp}
\usepackage{stfloats}
\usepackage{url}
\usepackage{verbatim}
\usepackage{graphicx}
\usepackage{xcolor}
\usepackage{multirow}
\def\BibTeX{{\rm B\kern-.05em{\sc i\kern-.025em b}\kern-.08em
    T\kern-.1667em\lower.7ex\hbox{E}\kern-.125emX}}
\usepackage{balance}
\begin{document}
\title{Mechanical ventilator development during COVID-19 crisis: Preclinical data analysis from porcine bio-model.\\ 
       \vspace{5mm}   
       \Large{Ventilador mecánico desarrollado durante la crisis del COVID-19: Análisis del estudio preclinico en biomoelos porcinos.}}
\author{Elian Conejo, Eduardo Calder\'{o}n, Carlos Araya, Ralph Garcia
\thanks{This works was supported by the program: ``Call for the Financing Program for Applied Research Projects and Technological Development, Incentive Fund Law 7169, National Challenges In Face of the COVID-19 Crisis'' form MICITT and the Vicerectoria de Investigaci\'{o}n of the Universidad de Costa Rica.}

\thanks{Elian Conejo, is with the Centro de Investigaci\'{o}n en Ciencias \'{A}tomicas, Nucleares y Moleculares (CICANUM), Universidad de Costa Rica, (e-mail: elian.conejo@ucr.ac.cr)}

\thanks{Eduardo Calder\'{o}n, is with the Escuela de Ingenier\'{i}a Mec\'{a}nica, Universidad de Costa Rica, (e-mail: eduardo.calderon@ucr.ac.cr ).}

\thanks{Carlos Araya is with the Escuela de Ingenier\'{i}a El\'{e}ctrica, Universidad de Costa Rica, (e-mail: carlos.arayajimenez@ucr.ac.cr).}

\thanks{Ralph Garcia is with the Escuela de F\'{i}sica, Universidad de Costa Rica, (e-mail: ralph.garcia@ucr.ac.cr ).}}

\markboth{Journal of \LaTeX\ Class Files,~Vol.~18, No.~9, September~2020}%
{How to Use the IEEEtran \LaTeX \ Templates}

\maketitle

\begin{abstract}
This paper describe a mechanical ventilator prototype with preclinical test performed on $10$ bioporcine models, where results have shown the capabilities to maintain physiological parameter for each subject under test and present also the capability of monitoring the pulmonary parameters, compliance ($C$), where this is the unique proposed prototype to present this capability at this extended subject samples. 
\end{abstract}

\begin{otherlanguage}{spanish} 
\begin{abstract}
Este articulo describe las pruebas realizadas de un ventilador mecánico en $10$ sujetos porcinos, donde los resultados muestran las capacidades del ventilador para sustentar y mantener los parámetros fisiológicos de cada sujeto bajo prueba, como también la capacidad de medir parámetros pulmonares como la complianza ($C$),  el cual seria el único prototipo con esta capacidad y único en ser probado en a esta cantidad de sujetos de prueba.
\end{abstract}
\end{otherlanguage}


\begin{otherlanguage}{spanish} 
\end{otherlanguage}

\section{Introduction}

\IEEEPARstart{T}he COVID-19, at its early stage of propagation, has put in evidence the requirements and necessities of the installed health facility capacities \cite{cheng2020,guan2020}. This situation has became a global health problem \cite{world2020coronavirus} attending the incremental number of COVID-19 cases. 

\par The acquired experiences dealing with the COVID-19 infection cases from patients with respiratory distress \cite{xu2020,gattinoni2020}, mechanical ventilation was the first procedure applied \cite{meng2020intubation,ng2020imaging} and recommended \cite{xie2020critical}.    

\par The COVID-19 effects to the respiratory system were the variation of the normal lungs biophysical parameters such as compliance, resistance and others \cite{pan2020}, resulting in a critical respiratory distress. For this reason the use of mechanical ventilation systems was vital at the early stages of this critical conditions, in order to help and support the patients recovery \cite{xie2020critical}. Thereafter, recommendations were submitted for the use of this medical devices to help the patients with this critical condition \cite{cook2020,meng2020,murthy2020}.

\par This paper describes a tested device of mechanical ventilator that have passed preclinical test, in order to increase the installed capacity for COVID-19 patients with respiratory distress, in case of shortage of mechanical ventilators, as it was taken place in different countries \cite{white2020,ranney2020} where the population was affected by the COVID-19.

\section{Principal aspect for an emergency mechanical ventilator}

\par Due to the shortage of mechanical ventilator as consequence of the COVID-19 pandemic, different mechanical ventilator designs were proposed during 2020. It is worthy to highlight the effort of the MHRA from United Kingdom \cite{RMVS001}, to issue guidelines of minimal clinically acceptable ventilator specifications, due to the potential shortage of ventilators supply. The HEV Ventilator Proposal has considered proposed prototypes, which can be quickly manufactured cheaply and on large scale is necessary, highlighting important technical aspects of the patient safety.

\par The aspects of the developed mechanical ventilator are describes in detail in Table \ref{tab:mechanical_ventilator_aspects}

\begin{table}[ht]
\small\sf\centering
\caption{Aspects of the mechanical ventilator.}
\begin{tabular}{l l l}
\hline
Parameter & Value & Observations \\
\hline\hline
Working Pressure & $0$ to $50 \hspace{1mm} cmH_2O$ &                 \\
\hline
Peak Pressure    &  $40 \hspace{1mm} cmH_2O$    &    Limited by \\
                 &                &    valve protection    \\
\hline
Operation Mode   &       VCV        &     \\
\hline
Flow             & $-100$ to $100 \hspace{1mm} L/min$ &                     \\
\hline
Inhalatory Trigger & $2$ to $6 \hspace{1mm} L/min$  & \\
\hline
Respiratory Rate & $10$ to $30$ cycles/min &                           \\
\hline
Inspiratory: & & \\
Expiratory ratio (I:E) &    1:5 to 5:1    &            \\
\hline
PEEP             & $5$ to $20 \hspace{1mm} cmH_2O$   & Regulated with \\
                 &                     & PEEP valve \\
\hline
Tidal Volume    & $100 \hspace{1mm} mL$ to $800 \hspace{1mm} mL$ & Depends on the \\
                &                      &  Ambu\textregistered Bag  \\
                &                      &  used                             \\
\hline\hline  

\end{tabular} 
\label{tab:mechanical_ventilator_aspects}

\end{table}

\section{Designing and development}

\subsection{Base pump system}

\par Due to the emergency at the pandemic situation, the use of disposable resuscitator (AmbuBag) as a pumping container was essential \cite{mora2020,hirani2020}, because this devise brings important advantage for a swift mechanical ventilator development process as fallow:
\begin{itemize}
	\item[a)] It is a tested and registered for medical uses device in almost all the countries.
	\item[b)] It is cheap and easy to acquired.
	\item[c)] It permits to get a controlled range volume of $100 \hspace{1mm} mL$ to $1000 \hspace{1mm}mL$.
	\item[d)] The PEEP valve and exhaust protection valve are easily implemented or its included in some models.
	\item[e)] It has compatibility with respiratory circuit tubing used in mostly all the hospitals.
	\item[f)] It has the adaptability to be connected to hospital oxygen outlets.    
\end{itemize}

\par Implementing this devices into the prototype developing process, in place of a compressor or turbine system, we were able to focus into developing electromechanical  and its respective sensor system.

\subsection{Electromechanical pump system}

\par Principally to delivery an oxygen mix volume using a disposable resuscitator, it must be deformed appropriately in a fixed frequency with an appropriate implemented control system. For this pumping process a linear electromechanical actuator was used.

\par Different actuator systems were proposed, such as NEMA motor based system among all the diffused mechanical ventilator prototype ideas. It is known that a NEMA motor well implemented with a feedback system, it could be very precises in motion and position, however without a validated reduction gearbox and encoder system, the prototype will not withstand at least for a few of continuous functioning hours.

\par For this propose, an industrial certificated linear actuator was chosen from FESTO, Germany. The principal parameters for an actuator for this application, pumping continuously an full oxygen mix AmbuBag are:

\begin{itemize}
	\item[a)] To have a position controlled range of $0$ to $100 \hspace{1mm} mm$.
	\item[b)] To have a minimum pushing force of $100 \hspace{1mm}N$.
	\item[c)] Long life cycles.
	\item[d)] Support continuous operation at least 50 days \cite{eimer2020}. 
\end{itemize}  

\subsection{Pressure and Flux sensor system}

\par
\begin{figure}[ht]
	\centering
	\includegraphics[width=8.8cm]{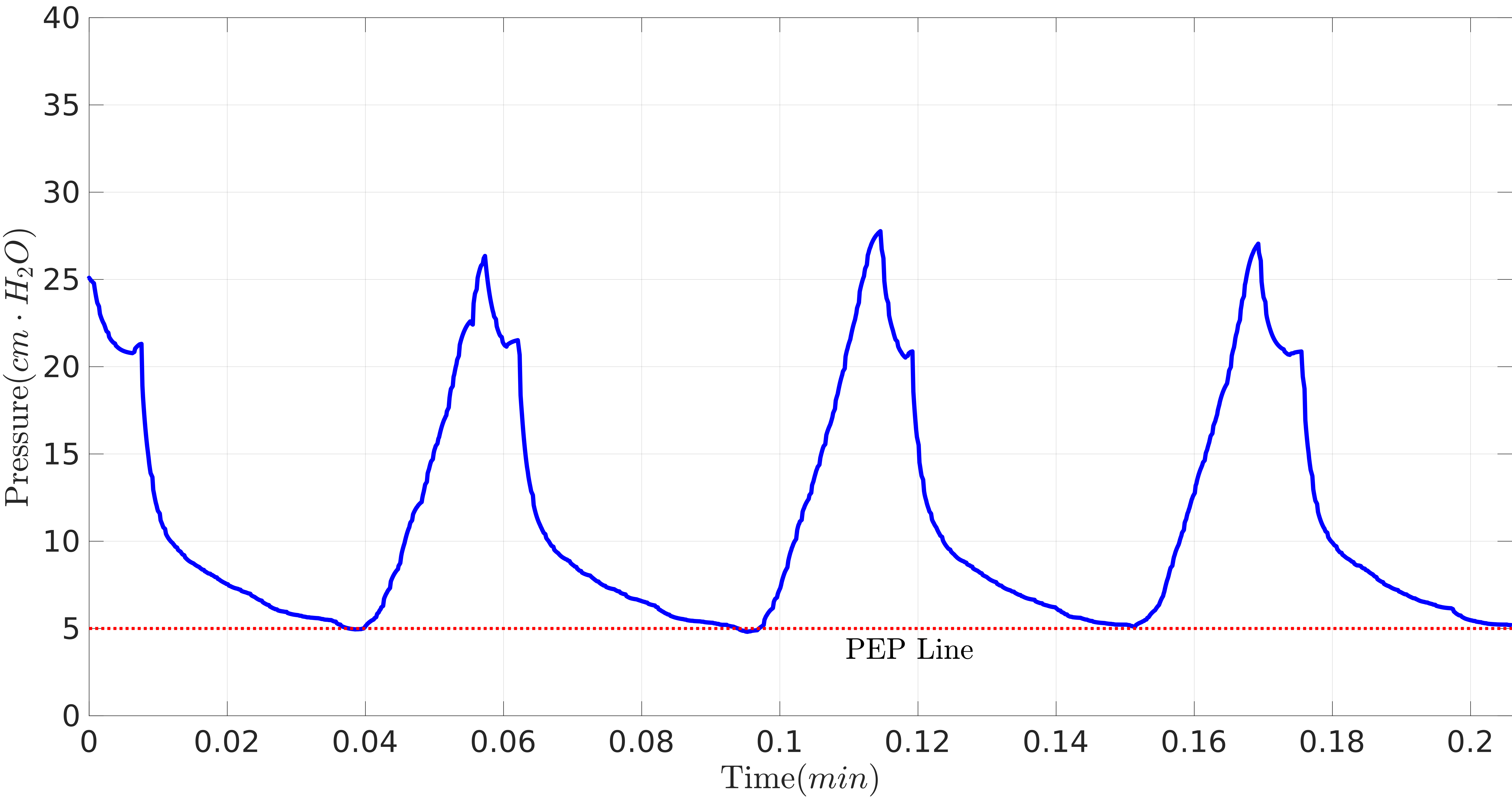}
	\caption{Prototype pressure signal from subject data analysis where the PEEP line remains constant over the subject intervention.}
	\label{fig:pressure_versus_time}
\end{figure}

 The pressure must be monitored as close as possible to the breathing circuit outlet. In this case the pressure is sampled on an outlet port of the filter. By measuring the pressure in real time, its enables us to plot this parameter on screen, visualizing the breathing cycle and setting the pressure range for the alarms: over pressure, under programmed PEEP pressure, breathing circuit leaks and other programming settings that are required. The pressure signal from the subject observation is depicted in Figure \ref{fig:pressure_versus_time}. 

\par
\begin{figure}[ht]
	\centering
	\includegraphics[width=8.8cm]{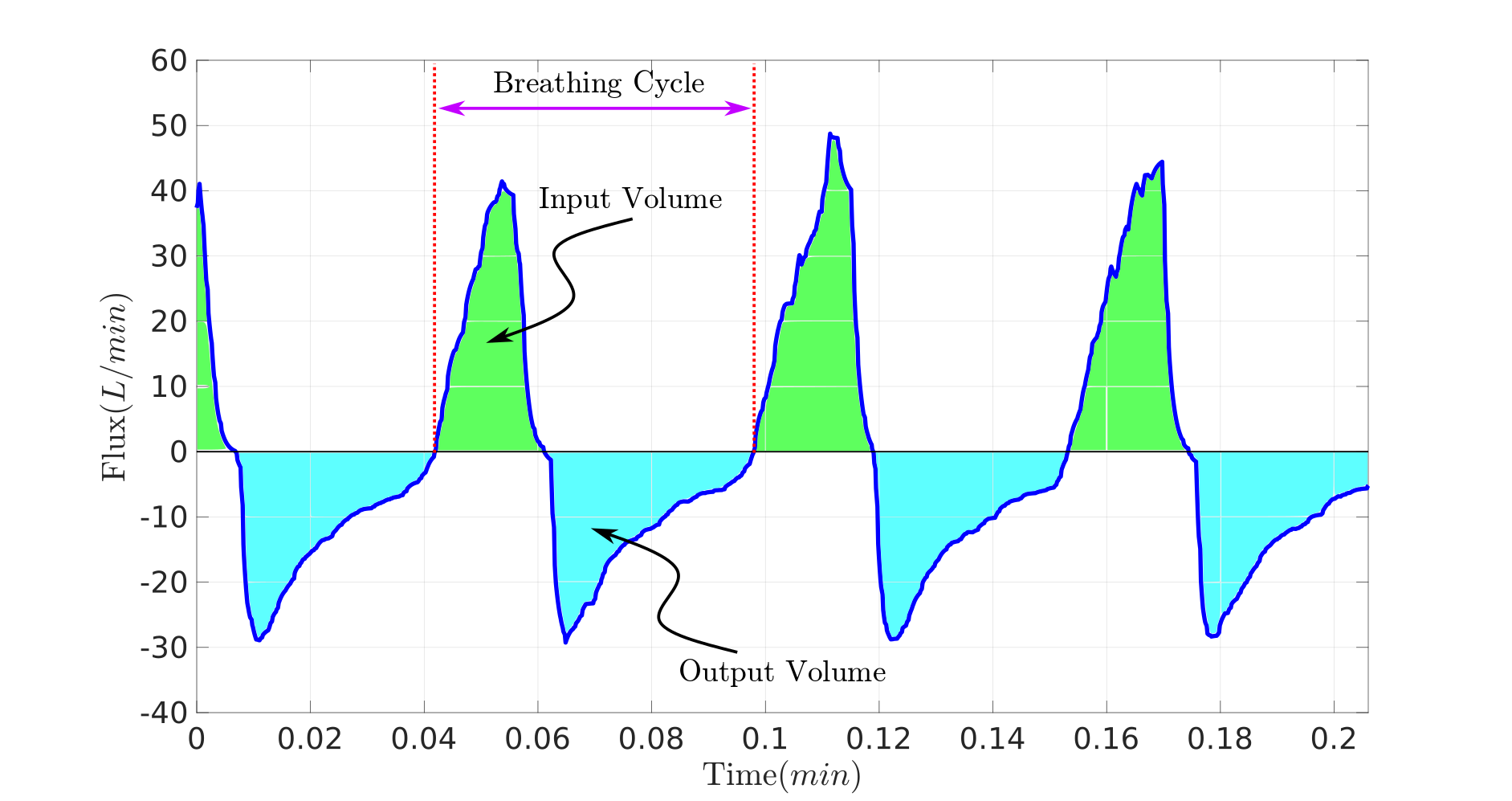}
	\caption{Prototype flux signal from subject data analysis, the volume can be computed by integration of the signal for each breathing cycle.}
	\label{fig:flux_to_volume_signal}
\end{figure}

 The flux is also measurement in real time. The sample rate of the flux measurement must be the high as possible and its signal well filtered in order to compute the delivered volume by performing an adequate integration procedure of the flux signal, as shown in Figure \ref{fig:flux_to_volume_signal}.

\subsection{Intuitive Graphical Interface}

\par
\begin{figure}[ht]
	\centering
	\includegraphics[width=8.8cm]{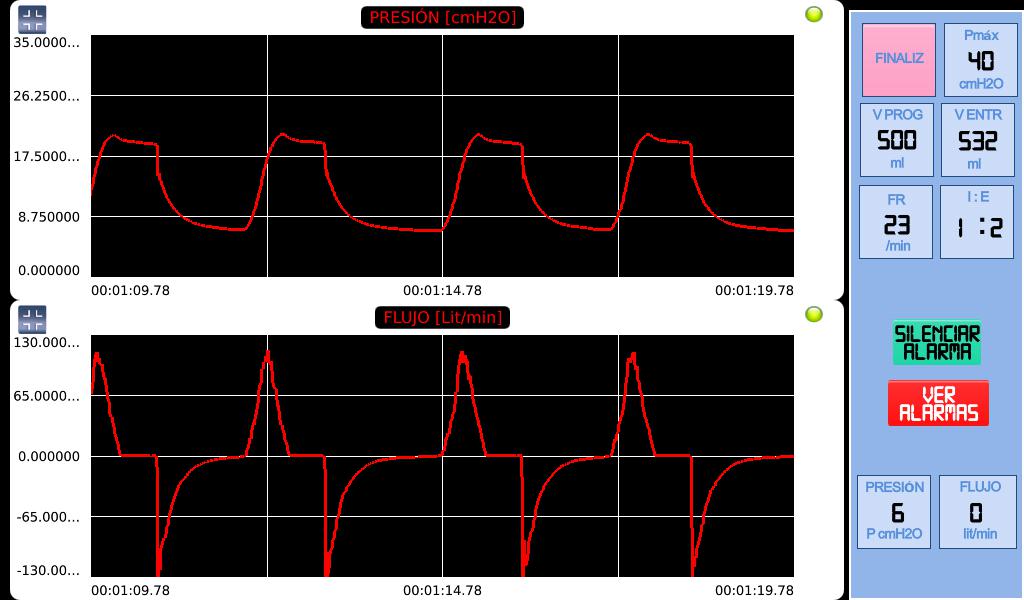}
	\caption{Mechanical ventilator screenshot of the touchscreen interface, where the pressure and tidal volume timeline is displayed and the setting parameter can be accessed.}
	\label{fig:ventilator_screenshot}
\end{figure}

\begin{figure}[ht]
	\centering
	\includegraphics[width=8.0cm]{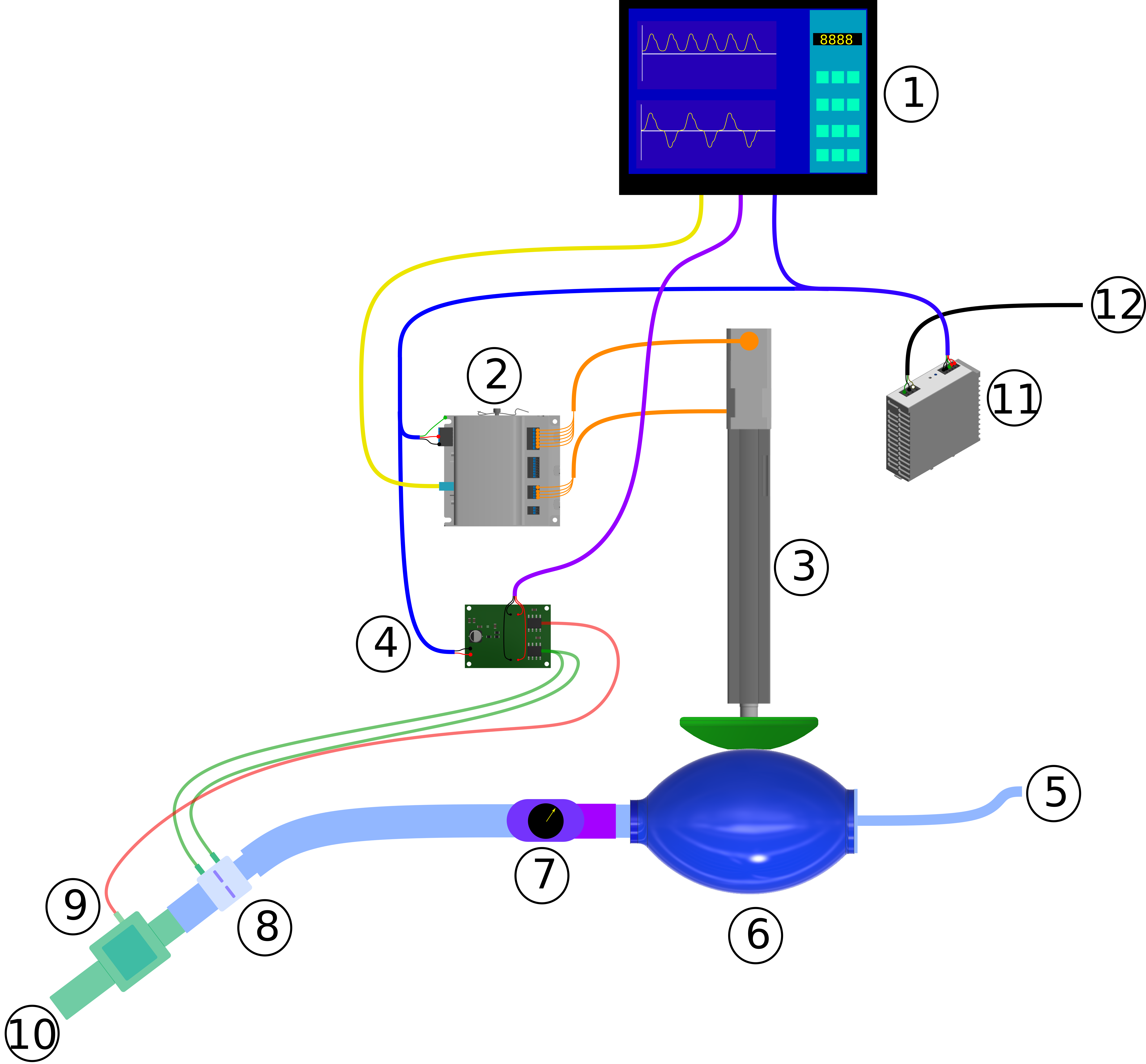}
	\caption{General scheme of the mechanical ventilator system, for numeric labels refer to Table \ref{tab:description_system}.}
	\label{fig:ventilator_system}
\end{figure}

 Above all the implementation part of the system, described in Figure \ref{fig:ventilator_system}, the graphical interface is an important component for visualizing, controlling and setting the working parameters for a mechanical ventilator. In our case, a touchscreen equipped with an electronic signal processing system was implemented in order to interact with the user and to display the settings, see Figure \ref{fig:ventilator_screenshot}. Furthermore, the touchscreen electronic system must have an ADC with an optimal sample rate to process the flux and pressure signals, performing the volume computation and displaying the signal on the screen. This condition must be revised on this kind of devices that are usually used on industrial applications and are cost effective, because not all the brands have the capability to implement an ADC with a sample rate more than $100$ samples per second.

\par For this reason, all the brands available on the market in Costa Rica, the US10-B10-TR22 touchscreen from UNITRONICS was implemented and its ADC is 12 bits of resolution with a sample rate of $100$. This enable us to get a smooth signal flux to compute the tidal volume accurately and an appropriated resolved pressure signal for constant monitoring and to set the alarm triggers properly.

\begin{table}[ht]
\small\sf\centering
\caption{Description of the Figure \ref{fig:ventilator_system}.}
\begin{tabular}{l l}
\hline
Label & Description \\
\hline
{1} & Touchscreen  \\ 
{2} & Linear Actuator Controller \\
{3} & Linear Actuator \\
{4} & Sensor Board \\
{5} & $O_2$ Mix Inlet \\
{6} & AmbuBag\textregistered\\
{7} & PEEP and Relief Valve \\
{8} & Flux Pitot Sensor and Check Valve \\  
{9} & Filter and Outlet Port for pressure measurement \\
{10} & Breathing Circuit Output \\
{11} & Power Supply $24 \hspace{1mm} V$ DC @ $5.0 \hspace{1mm} Amp$ \\
{12} & To Main $220/110 \hspace{1mm} V$ AC \\
\hline \hline

\end{tabular}\label{tab:description_system}

\end{table}

\section{Pre-clinical test on porcine bio-model and Results}

\par The porcine bio-model has been the milestone stage in biomedical research for many years. In medical respiratory research field, the size and biophysical characteristics of porcine lungs are similar to human lungs system, giving the best condition, as a bio-model, for testing and optimizing a mechanical ventilation device \cite{judge2014}.

\par For this research project, genetic line pigs of low predisposition of respiratory illness were used, from an animal lot with sanitary traceability profile in order to assuring subjects free from any respiratory problems. The animal lot was composed by seven females from Topig genetic line, three months age old, mass range between $25$ and $30$ kilogram, risen and moved to laboratory under the highest animal welfare conditions.          

\par The subjects were placed in vivarium $72$ hours before clinical procedure, where an objective structured clinical examination (OSCE) procedure was applied to verify and validate the optimal health conditions. The management of the subjects in vivarium was performed in agreement with international welfare animal compliance for this specie (NRC) \cite{national2010}, such procedure was verified and approved by Institutional Animal Care and Use Committee (IACUC) of the Universidad de Costa Rica. 

\par At the beginning of clinical procedure, the subject were anesthetized using TIVA procedure with midazolam and ketamine for initial procedural sedation, in combination with midazolam-ketamine-propofol sedation during the interventional procedure. The rocuronium was used for neuromuscular blocking of the subject into the interventional procedure.   

\par
\begin{figure}[ht]
	\centering
	\includegraphics[width=8.0cm]{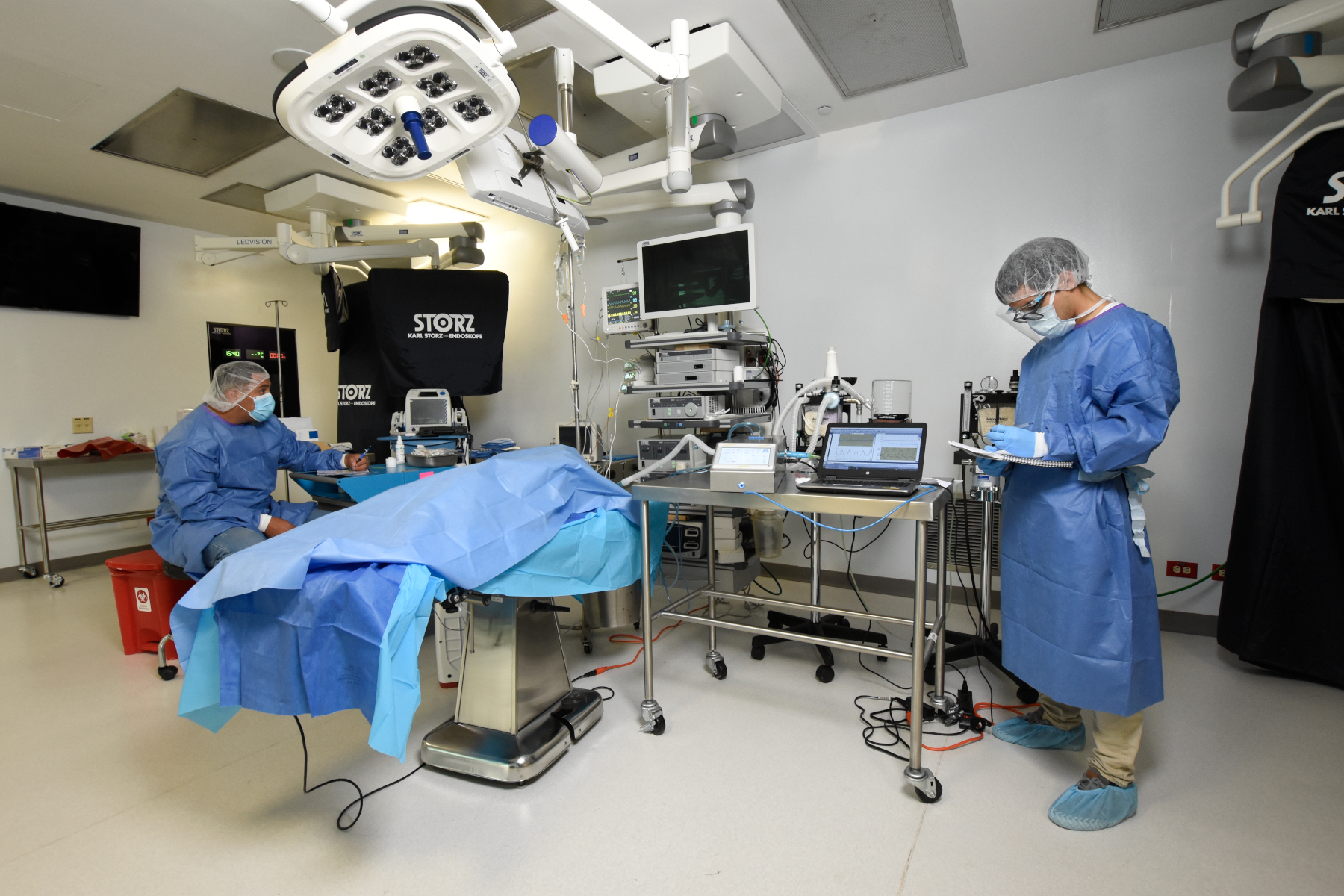}
	\caption{Subject under care during the preclinical ventilator test. The mechanical ventilator was connected to a PC for recording data in real time for further analysis.}
	\label{fig:preclinical_procedure_in_vivo}
\end{figure}

 The tracheal intubation procedure was applied for every subject, following real time monitoring of $O_2$ saturation, electrocardiogram, heart rate, breath rate, non invasive arterial pressure and corporeal temperature. These parameters were evaluated and recorded every $10$ minutes. The subjects were hydrated properly and keeping warmed with electric blanket as shown in Figure \ref{fig:preclinical_procedure_in_vivo}.

\par The overall time taken for the preclinical test for each subject was $7$ hours, applied to $10$ subjects.
 
\par The mechanical ventilators were consecutive submitted to several tests on the subjects for a determined period of time, following a designed experimental clinical procedure, simulating all possible situations that could be shown in a human patience needed of this medical devices. For each test, the subjects were exposed to a different degree of breathing restrictions, in order to verify the performance of the mechanical ventilator to restore and maintain the breathing parameters.

\par Specifically, the performed test were: mechanical ventilation without breathing restriction, with neuromuscular blocking, neuromuscular blocking with hyperoxygenation, neuromuscular blocking with hyperventilation, induced moderated respiratory distress and induced severe respiratory distress. At the final stage of the clinical procedure, each subject were humanely euthanized and a necropsy procedure were performed to verify any injury due to external factors.
     
\par
\begin{figure}[ht]
	\centering
	\includegraphics[width=8.8cm]{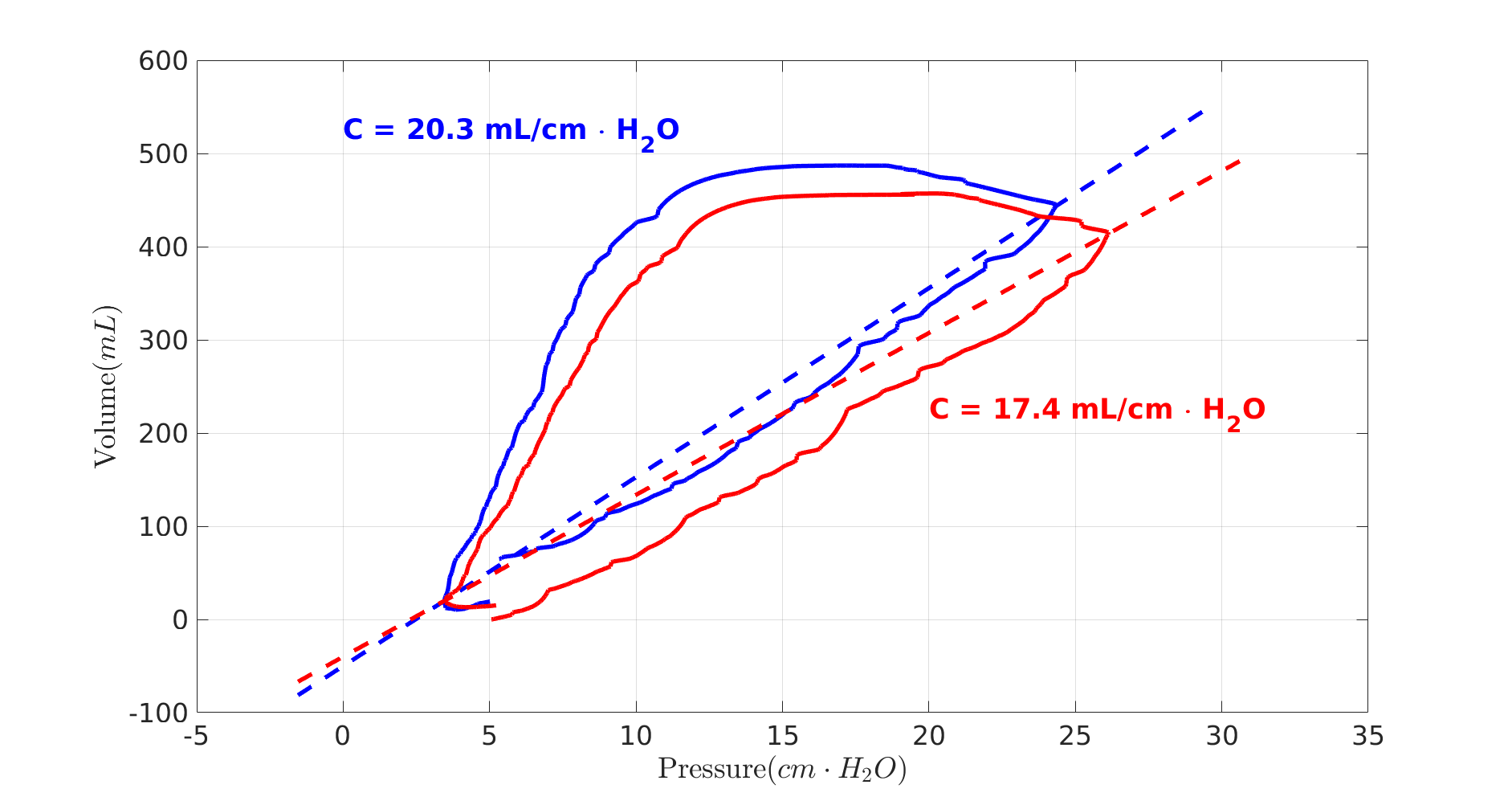}
	\caption{Pressure-Volume hysteresis loop from subject data before (\textcolor{blue}{- - -}) and after (\textcolor{red}{- - -}) removing pulmonary surfactant.}
	\label{fig:pressure_volume_loop}
\end{figure}

 An induced severe respiratory distress by removing the pulmonary surfactant was the final stage of the preclinical test. The recorded data from the mechanical ventilator was possible to verify the behavior of the lungs by plotting Volume-Pressure Loop, seen in Figure \ref{fig:pressure_volume_loop}. 
 
\par Finally, the results computed from plotted data in Figure \ref{fig:pressure_volume_loop} are described in Table \ref{tab:computed_data}, where the evolution of pulmonary parameters of the subject, under those two detailed conditions, are quantified.   

\begin{table}[ht]
\scriptsize\centering
\caption{Pulmonary parameters computed from Pressure-Volume hysteresis loop data recorded by the mechanical ventilator under test.}
\begin{tabular}{c l r c}
\cline{2-4}
 & Parameter & Values & Units \\
\hline
\multirow{3}{*}{Initial State} & Compliance & $20.3$ & $mL/cmH_2O$ \\
                               & Hysteresis Loop Area & $4463.44$ & $mL\cdot cmH_2O$ \\
\hline

\multirow{3}{*}{Final State} & Compliance & $17.4$ & $mL/cmH_2O$ \\
                               & Hysteresis Loop Area & $4933.73$ & $mL\cdot cmH_2O$ \\
\hline

\multirow{3}{*}{Relative Variations} & $\Delta$ Compliance & $-14.3$ & $\%$ \\
                               & $\Delta$Hysteresis Loop Area & $10.5$ & $\%$ \\
\hline
\end{tabular}\label{tab:computed_data}
\end{table}  
 
\par
\begin{figure}[ht]
	\centering
	\includegraphics[width=8.8cm]{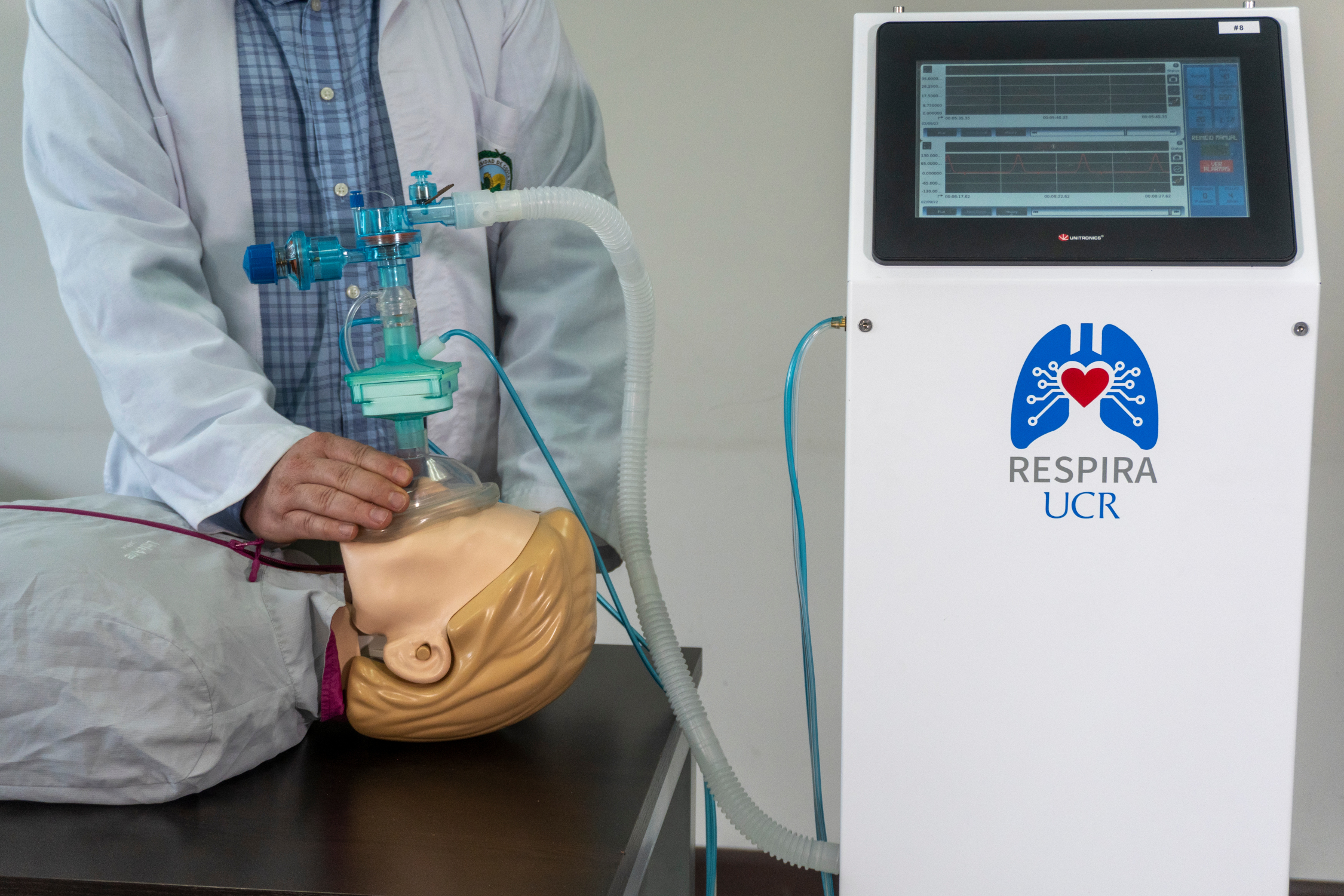}
	\caption{Prototype under demonstration with the final-designed case, ready to be implemented in an ICU or primal care facility in case of shortage of mechanical ventilation units.}
	\label{fig:final_case_prototype}
\end{figure}

 Here we can see, in Table \ref{tab:computed_data}, the pulmonary compliance $dV/dP$ decreases $14.3\%$ due to removing procedure of the pulmonary surfactant at the last stage of induce severe respiratory distress procedure. All the subject, at this stage of the preclinical procedure, were supported efficiently by the mechanical ventilator prototype. The Figure \ref{fig:final_case_prototype} shown the final prototype with the designed requirements to be implemented in a ICU or primary care facility in case of a shortage situation of mechanical ventilation units. 
 
\section{Conclusions}

\par The preclinical analysis was applied to $10$ subjects, where the performance of the prototype was verified by evaluating that each subject physiological parameter remains and returns swiftly to the normal values for each induced stress, regarding the preclinical protocol. 

\par One remarkable result was the prototype is capable to produce data for evaluating the evolution of pulmonary parameters of the subject under test, as is shown in Figure \ref{fig:pressure_volume_loop}, where this is the unique proposed prototype to present this capability.

\par At the final stage of the preclinical protocol, where induced stress by removing pulmonary surfactant, the prototype was successful to support the subject under test in this induced state, close to a critical respiratory distress.

\section*{Acknowledgment}

\par The authors would like to acknowledge Dr. Steven Quirós and Dr. Luis Zuñiga for their contribution using the DCLab, now CICICA, facilities and their expertise for the preclinical test and Dr. Jilma Alemán, Dr. Gilberth Alvarado, Dr. Denis Chavez, Dr. Kwok Ho Sánchez, Dr. Alberto Cubero and Dr. Daniel Vindas for your guidance and support in this project. Dr, Jaime Caravaca M. for making possible the simulation test on the CECISA-UCR. We are grateful for ROCHE’s logistical support in order to carry out the preclinical test, Fedefarma for its contribution of this process, Cámara Costarricense de Porcicultores specially Mr. DeVitre for the subjects samples. We want to thanks Elvatron for its contribution on electronics parts supply, the diplomatic corp from Germany, Korea, China and Switzerland for their contribution to this project and the Rectoria, the Vicerrectoria de Investigación of the Universidad de Costa Rica for its constant support and the Ministry of Science, Technology and Telecommunications of Costa Rica. Finally, we want to thank Mr. Victor Rodriguez to construct some of the mechanical pieces of the first prototype.

\bibliographystyle{IEEEtran}
\bibliography{mybibliography.bib}

\begin{thebibliography}{10}
\providecommand{\url}[1]{#1}
\csname url@samestyle\endcsname
\providecommand{\newblock}{\relax}
\providecommand{\bibinfo}[2]{#2}
\providecommand{\BIBentrySTDinterwordspacing}{\spaceskip=0pt\relax}
\providecommand{\BIBentryALTinterwordstretchfactor}{4}
\providecommand{\BIBentryALTinterwordspacing}{\spaceskip=\fontdimen2\font plus
\BIBentryALTinterwordstretchfactor\fontdimen3\font minus
  \fontdimen4\font\relax}
\providecommand{\BIBforeignlanguage}[2]{{%
\expandafter\ifx\csname l@#1\endcsname\relax
\typeout{** WARNING: IEEEtran.bst: No hyphenation pattern has been}%
\typeout{** loaded for the language `#1'. Using the pattern for}%
\typeout{** the default language instead.}%
\else
\language=\csname l@#1\endcsname
\fi
#2}}
\providecommand{\BIBdecl}{\relax}
\BIBdecl

\bibitem{cheng2020}
V.~C. Cheng, S.-C. Wong, J.~H. Chen, C.~C. Yip, V.~W. Chuang, O.~T. Tsang,
  S.~Sridhar, J.~F. Chan, P.-L. Ho, and K.-Y. Yuen, ``Escalating infection
  control response to the rapidly evolving epidemiology of the coronavirus
  disease 2019 (covid-19) due to sars-cov-2 in hong kong,'' \emph{Infection
  Control \& Hospital Epidemiology}, vol.~41, no.~5, pp. 493--498, 2020.

\bibitem{guan2020}
W.-j. Guan, W.-h. Liang, Y.~Zhao, H.-r. Liang, Z.-s. Chen, Y.-m. Li, X.-q. Liu,
  R.-c. Chen, C.-l. Tang, T.~Wang \emph{et~al.}, ``Comorbidity and its impact
  on 1590 patients with covid-19 in china: a nationwide analysis,''
  \emph{European Respiratory Journal}, vol.~55, no.~5, 2020.

\bibitem{world2020coronavirus}
W.~H. Organization \emph{et~al.}, ``Coronavirus disease 2019 (covid-19):
  situation report, 73,'' 2020.

\bibitem{xu2020}
Z.~Xu, L.~Shi, Y.~Wang, J.~Zhang, L.~Huang, C.~Zhang, S.~Liu, P.~Zhao, H.~Liu,
  L.~Zhu \emph{et~al.}, ``Pathological findings of covid-19 associated with
  acute respiratory distress syndrome,'' \emph{The Lancet respiratory
  medicine}, vol.~8, no.~4, pp. 420--422, 2020.

\bibitem{gattinoni2020}
L.~Gattinoni, S.~Coppola, M.~Cressoni, M.~Busana, S.~Rossi, and D.~Chiumello,
  ``Covid-19 does not lead to a “typical” acute respiratory distress
  syndrome,'' \emph{American journal of respiratory and critical care
  medicine}, vol. 201, no.~10, pp. 1299--1300, 2020.

\bibitem{meng2020intubation}
L.~Meng, H.~Qiu, L.~Wan, Y.~Ai, Z.~Xue, Q.~Guo, R.~Deshpande, L.~Zhang,
  J.~Meng, C.~Tong \emph{et~al.}, ``Intubation and ventilation amid the
  covid-19 outbreak: Wuhan’s experience,'' \emph{Anesthesiology}, vol. 132,
  no.~6, pp. 1317--1332, 2020.

\bibitem{ng2020imaging}
M.-Y. Ng, E.~Y. Lee, J.~Yang, F.~Yang, X.~Li, H.~Wang, M.~M.-s. Lui, C.~S.-Y.
  Lo, B.~Leung, P.-L. Khong \emph{et~al.}, ``Imaging profile of the covid-19
  infection: radiologic findings and literature review,'' \emph{Radiology:
  Cardiothoracic Imaging}, vol.~2, no.~1, p. e200034, 2020.

\bibitem{xie2020critical}
J.~Xie, Z.~Tong, X.~Guan, B.~Du, H.~Qiu, and A.~S. Slutsky, ``Critical care
  crisis and some recommendations during the covid-19 epidemic in china,''
  \emph{Intensive care medicine}, vol.~46, no.~5, pp. 837--840, 2020.

\bibitem{pan2020}
F.~Pan, T.~Ye, P.~Sun, S.~Gui, B.~Liang, L.~Li, D.~Zheng, J.~Wang, R.~L.
  Hesketh, L.~Yang \emph{et~al.}, ``Time course of lung changes at chest ct
  during recovery from coronavirus disease 2019 (covid-19),'' \emph{Radiology},
  vol. 295, no.~3, pp. 715--721, 2020.

\bibitem{cook2020}
T.~Cook, K.~El-Boghdadly, B.~McGuire, A.~McNarry, A.~Patel, and A.~Higgs,
  ``Consensus guidelines for managing the airway in patients with covid-19:
  Guidelines from the difficult airway society, the association of
  anaesthetists the intensive care society, the faculty of intensive care
  medicine and the royal college of anaesthetists,'' \emph{Anaesthesia},
  vol.~75, no.~6, pp. 785--799, 2020.

\bibitem{meng2020}
L.~Meng, H.~Qiu, L.~Wan, Y.~Ai, Z.~Xue, Q.~Guo, R.~Deshpande, L.~Zhang,
  J.~Meng, C.~Tong \emph{et~al.}, ``Intubation and ventilation amid the
  covid-19 outbreak: Wuhan’s experience,'' \emph{Anesthesiology}, vol. 132,
  no.~6, pp. 1317--1332, 2020.

\bibitem{murthy2020}
S.~Murthy, C.~D. Gomersall, and R.~A. Fowler, ``Care for critically ill
  patients with covid-19,'' \emph{Jama}, vol. 323, no.~15, pp. 1499--1500,
  2020.

\bibitem{white2020}
D.~B. White and B.~Lo, ``A framework for rationing ventilators and critical
  care beds during the covid-19 pandemic,'' \emph{Jama}, vol. 323, no.~18, pp.
  1773--1774, 2020.

\bibitem{ranney2020}
M.~L. Ranney, V.~Griffeth, and A.~K. Jha, ``Critical supply shortages—the
  need for ventilators and personal protective equipment during the covid-19
  pandemic,'' \emph{New England Journal of Medicine}, vol. 382, no.~18, p. e41,
  2020.

\bibitem{RMVS001}
MHRA-UK, ``Rapidly manufactured ventilator system,'' \emph{Medicine \&
  Healthcare products Regulatory Agency, UK}, 2020.

\bibitem{mora2020}
S.~Mora, F.~Duarte, and C.~Ratti, ``Can open source hardware mechanical
  ventilator (osh-mvs) initiatives help cope with the covid-19 health crisis?
  taxonomy and state of the art,'' \emph{HardwareX}, p. e00150, 2020.

\bibitem{hirani2020}
H.~Hirani, ``Mechanical ventilator using motorized bellow,'' \emph{Transactions
  of the Indian National Academy of Engineering}, vol.~5, no.~2, pp. 379--384,
  2020.

\bibitem{eimer2020}
J.~Eimer, J.~Vesterbacka, A.-K. Svensson, B.~Stojanovic, C.~Wagrell,
  A.~S{\"o}nnerborg, and P.~Nowak, ``Tocilizumab shortens time on mechanical
  ventilation and length of hospital stay in patients with severe covid-19: a
  retrospective cohort study,'' \emph{Journal of internal medicine}, 2020.

\bibitem{judge2014}
E.~P. Judge, J.~L. Hughes, J.~J. Egan, M.~Maguire, E.~L. Molloy, and
  S.~O’Dea, ``Anatomy and bronchoscopy of the porcine lung. a model for
  translational respiratory medicine,'' \emph{American journal of respiratory
  cell and molecular biology}, vol.~51, no.~3, pp. 334--343, 2014.

\bibitem{national2010}
N.~R. Council \emph{et~al.}, ``Guide for the care and use of laboratory
  animals,'' 2010.

\end{thebibliography}

\end{document}